\documentclass[12pt,preprint]{aastex}

\def\chandra {{\it Chandra}}
\def\kms{\rm\,km\,s^{-1}}
\def\ergcms{\rm\,erg\,cm^{-2}\,s^{-1}}
\def\l5100{L_{\rm 5100\AA}}
\def\l1kev{L_{\rm 1 keV}}
\def\mbh{M_{\rm BH}}
\begin{document}

\title{\chandra\ Observations of X-Ray Weak Narrow-Line Seyfert 1 Galaxies}
\author{Rik J. Williams, Smita Mathur, and Richard W. Pogge}
\affil{Department of Astronomy, Ohio State University, 140 West 18th Avenue, Columbus, OH 43210-1173, USA}
\email{Send comments to: williams@astronomy.ohio-state.edu}
\shorttitle{X-Ray Weak Narrow-Line Seyfert 1 Galaxies}
\shortauthors{Williams, Mathur, \& Pogge}

\begin{abstract}

We present \chandra\ observations of 17 optically-selected,
X-ray weak narrow-line Seyfert 1 (NLS1) galaxies.  These objects were
optically identified by Williams et al.~(2002) in the Sloan Digital Sky
Survey Early Data Release, 
but were not found in the ROSAT All-Sky Survey (RASS) despite having
optical properties similar to RASS--detected NLS1s.  All objects in this
sample were detected by \chandra\ and exhibit a range of 0.5--2~keV 
photon indices
$\Gamma=1.1-3.4$.  One object was not detected in the soft band, but has
a best--fit $\Gamma=0.25$ over the full 0.5--8~keV range.  These
photon indices extend to values far below what are normally observed in NLS1s. 
A composite X-ray spectrum of the hardest objects in this sample
does not show any signs of absorption, although the data do not prohibit
one or two of the
objects from being highly absorbed.  We also find a strong 
correlation between
$\Gamma$ and $\l1kev$; this may be due to differences in
$L_{\rm bol}/L_{\rm Edd}$ among the NLS1s in this sample.  Such variations
are seemingly in conflict with the current paradigm that NLS1s accrete
near the Eddington limit.  Most importantly,
this sample shows that strong, ultrasoft X-ray emission is not a universal
characteristic of NLS1s; in fact, a substantial number may exhibit weak
and/or low--$\Gamma$ X-ray emission.

\end{abstract}


\section{Introduction}

\citet{ostpog85} initially defined narrow-line Seyfert 1 galaxies
 (NLS1s) by their striking optical spectral characteristics: strong,
 narrow H$\beta$ emission \citep[later formally defined to be FWHM~$\leq
 2000 \kms$ by][]{goodrich89}, weak [\ion{O}{3}] relative to H$\beta$,
 and strong \ion{Fe}{2}.  These properties put NLS1s at one extreme end
 of the so--called \citet{bg92} ``Eigenvector 1,'' thought to correspond
 to emission from lower-mass nuclear black holes coupled with
 near--Eddington accretion rates \citep{boroson02}.  X-ray observations
 have revealed a strong soft X-ray ``excess'' in NLS1s
 \citep[e.g.,][]{leighly99}, further bolstering the high $L_{\rm bol}/L_{\rm
 Edd}$---low--mass black hole hypothesis \citep{pounds95,wang96}.
 Indeed, \citet[][hereafter BBF96]{bbf96} found a possible anticorrelation
 between X-ray spectral slope and H$\beta$ line width, with NLS1s generally
 having softer X-ray spectra than other AGN.  Ultrasoft X-ray
 selection has consequently proven to be an essential tool for the
 discovery of large numbers of NLS1s \citep[e.g.][]{grupe00,grupe04}.

The disadvantage of selecting NLS1s solely upon their X-ray
properties is that it can introduce into NLS1 samples a strong bias toward
those exhibiting an ultrasoft excess \citep{forster99}.  
Since NLS1s are primarily
defined by their optical properties, the true nature of their
X-ray emission is thus difficult to determine.  Though it is well known
that \emph{some} NLS1s are ultrasoft X-ray sources, previous samples of
optically selected NLS1s were simply too sparse to determine how many,
as well as whether or not a significant number of NLS1s have hard X-ray
spectra.  With the advent of the Sloan Digital Sky Survey
\citep[SDSS;][]{york2000}, it is possible to build large catalogues of
NLS1s with homogeneous selection criteria based on their optical spectra
alone \citep[see][hereafter WPM02]{wpm02}.  

Of the 150 NLS1s listed in WPM02, 52 were detected in the ROSAT All-Sky
Survey \citep[RASS;][]{rass}.  Forty--five of these had sufficient
counts in the 0.1--2.0~keV range to derive power-law photon indices
($\Gamma$, where $N(E)\propto E^{-\Gamma}$) based on hardness ratios.
Most of these objects were optically bright ($g \la 18.5$),
low--redshift ($z \la 0.4$), and exhibited typical NLS1 photon indices
of $\Gamma \ga 2.0$.  However, a substantial number of
optically--bright, low--redshift NLS1s did \emph{not} have RASS source
identifications.  The optical spectra of these objects appear completely
normal (within the limitations of the SDSS resolution and
signal--to--noise) and only one or two are in regions of high Galactic
\ion{H}{1} column density, which could potentially obscure the X-ray
flux.

It is thus possible that these objects represent a subset of NLS1s which are
optically normal but X-ray weak.  Only a few such objects have
previously been found; for example, 
RX J2217.9-5941 \citep{grupe01b} and PHL 1811 \citep{leighly01}. 
Since these SDSS NLS1s were not detected in 
the RASS, however, the nature and extent of their X-ray weakness could not be 
determined.  They could emit significant flux at
higher energies not covered by the 0.1--2.4~keV ROSAT band, or they
might also be ultrasoft X-ray sources but with substantially lower
overall X-ray fluxes than typically seen in NLS1s (i.e., much higher
$\alpha_{\rm{ox}}$).  Another possible explanation is variability, but
it seems unlikely that such a large fraction ($\sim 40\%$) would be in
an exceptionally low state during the RASS observations.  In reality,
all three of these factors probably have some bearing on the X-ray
weakness of these objects, but we cannot infer how many are affected by
which factors, if any, from the existing data.

As a first step toward solving this puzzle, we have observed 17 of these
optically--selected but RASS--undetected NLS1s with \chandra.  Due to its
excellent sensitivity and low background levels, \chandra\ is able to
detect objects at far lower flux levels than ROSAT, and its large
(0.5--8~keV) energy range allows detection of objects with harder X-ray
spectra as well.  Our primary goals are (1) to detect these objects in
X-rays, or set upper limits to their X-ray emission, and (2) to obtain
rough estimates of $\Gamma$ for the \chandra--detected NLS1s.  Given
this information, we can gain some insight as to which of the
aforementioned scenarios (if any) sufficiently explain these
ROSAT--unobserved NLS1s.  In this paper we present the results of these
\chandra\ observations, and the possible implications for the NLS1s in
our sample.

\section{Target Selection}

A full description of the spectroscopic selection and preliminary X-ray
analysis can be found in WPM02; a brief summary follows.  The NLS1s in
the WPM02 sample were initially selected from the SDSS Early Data
Release \citep{stoughton02} solely on the basis of narrow H$\beta$
emission.  Subsequently, each spectrum was visually inspected and a more
accurate measurement was taken of the H$\beta$ line width.  Objects
which did not fulfill the criteria of \citet{ostpog85} and
\citet{goodrich89} were discarded.  X-ray power law slopes and
luminosities were estimated for the 45 objects with adequate RASS data
available.

\begin{figure}
\plotone{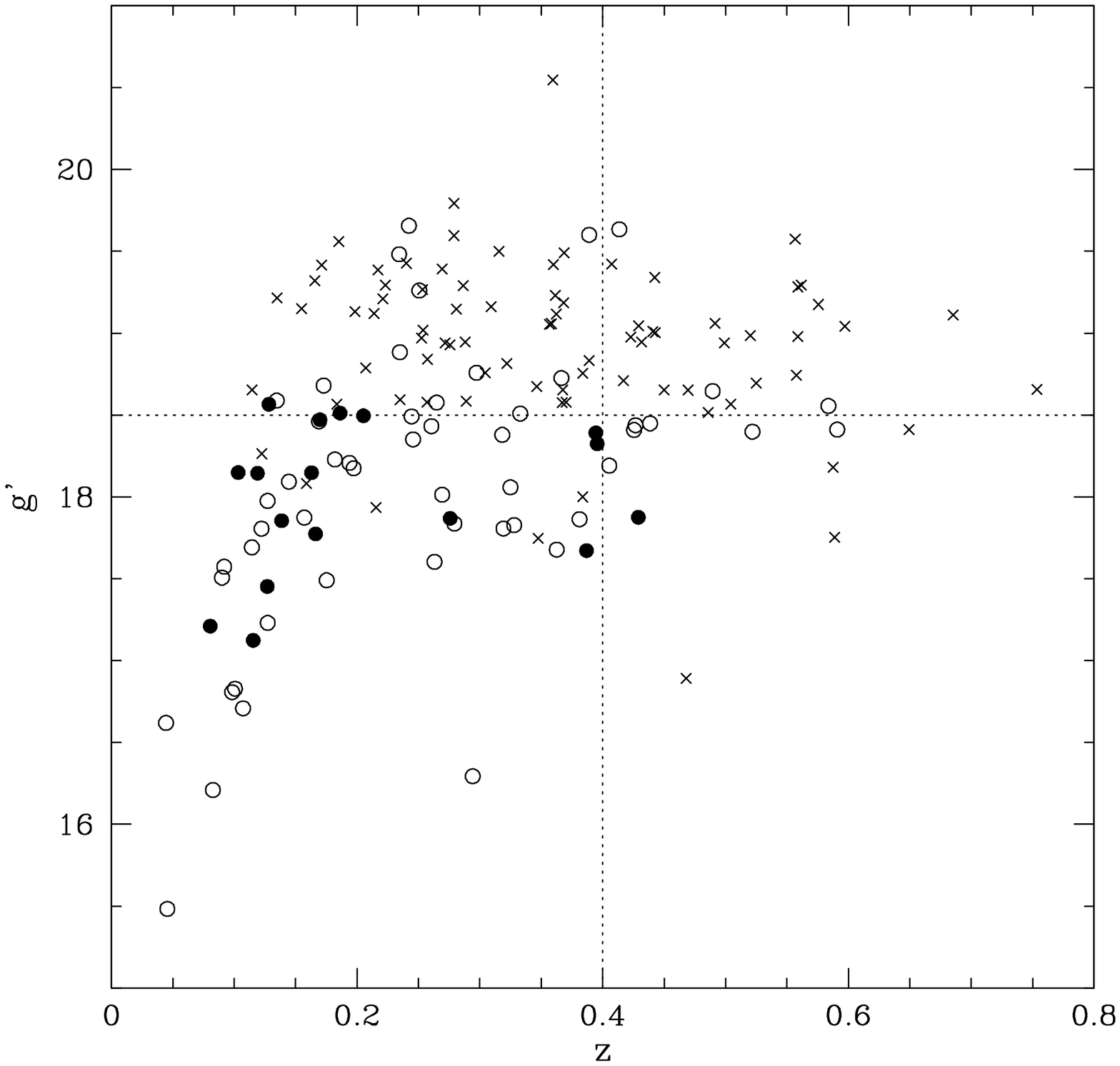}
\caption{SDSS g$^\prime$ magnitude--redshift relation for the WPM02 
SDSS--selected NLS1s.
Crosses indicate objects undetected in the RASS, open circles are 
RASS--detected objects,
and filled circles are NLS1s undetected in the RASS which were
 chosen for \chandra\ follow--up.
The dashed lines show the nominal brightness and redshift limits for 
the \chandra\ follow--up sample. \label{fig_gz}}
\end{figure}

Figure~\ref{fig_gz} shows the SDSS $g$~magnitude--redshift distribution
for the WPM02 sample, differentiating between NLS1s detected in RASS,
those not detected, and those chosen for \chandra\ follow-up.  In this
diagram, the fainter, higher--redshift NLS1s are detected in the RASS
less frequently, and most NLS1s with $g \la 18.5$ and $z \la 0.4$ were
detected by RASS.  We adopted these as rough limits for our follow--up
sample.  To rule out the possibility that high foreground extinction
prevented some objects from being detected in the RASS, we restricted
our sample to objects in regions of low Galactic \ion{H}{1} column
density\footnote{Found for each location using the {\tt nh} utility,
part of the HEAsoft package, available at
\url{http://heasarc.gsfc.nasa.gov/lheasoft/}.}  ($N_{\rm H} < 4\times
10^{20}$~cm$^{-2}$, although one relatively bright NLS1 in the sample
has $N_{\rm H} = 5.7\times 10^{20}$~cm$^{-2}$).  The resulting 17
objects which comprise this follow--up sample appear normal in all other
respects.


\section{Observations and Data Reduction}

We observed the 17 NLS1s in this sample with \chandra\ for $\sim 2$~ksec
each between 3 January and 15 September 2003.  The Advanced CCD Imaging
Spectrometer Spectroscopic array \citep[ACIS-S;][]{garmire03} was
employed, and all observations were offset by 3\arcmin\ from the nominal
aimpoint along each detector axis to mitigate the effects of pileup
should any of the sources be unexpectedly bright.  This resulted in a
slight ($\sim 10\%$) reduction in detector efficiency as well as a
broadening of the point--spread function.  Neither of these effects
significantly hinders our source detection efficiency, thanks to
\chandra's exceptionally low background levels.  These observing
parameters allow $3\sigma$ detections of point sources down to
$2-5\times 10^{-14}\ergcms$ (for $\Gamma = 3.0-1.0$ respectively) in the
0.5--8 keV band.

\subsection{Data preparation}
All data were reprocessed with the newest version of the \chandra\
Interactive Analysis of Observations (CIAO 3.0.1) software employing the
most recent calibration files (CALDB version 2.23).\footnote{Both
available at \url{http://cxc.harvard.edu/ciao/download\_ciao\_reg.html}}
We then defined circular extraction regions that were centered on each
source and of sufficient radius to encompass the PSFs (typically about
4\farcs 5).  Background regions were defined as annuli with inner and
outer radii of 7\farcs 5 and 15\arcsec\ respectively.  The CIAO tool
{\tt psextract} was used to extract source and background spectra for
those objects with sufficient flux ($\ga 25$ total counts).
Additionally, we used {\tt dmextract} to determine the raw number of
counts for each source in the 0.5--0.9~(S), 0.9--2.0~(M), and
2.0--8.0~keV~(H) bands.  Finally, the {\tt apply\_acisabs} script was
applied to each ancillary response file (ARF) in order to account for
the time--dependent ACIS quantum efficiency degradation.\footnote{See
\url{http://cxc.harvard.edu/cal/Acis/Cal\_prods/qeDeg/} for details.}

All 17 NLS1s in this sample were detected by \chandra, 16 with high
($\geq 4\sigma$) significance.  Table~\ref{tab_obslog} lists the net,
background subtracted, number of counts per energy bin for each object.
As expected, the background was found to be low, with 0--2 counts per
band detected in the background extraction regions (i.e. less than 0.25
counts expected in the source region on average), for most of the
observations. In only three cases were one or more background counts 
expected in the source region. Due to the low background levels, we 
assume the errors are simply Poisson errors on the raw number of 
source counts.

\subsection{Determination of Power-Law Slope}
We used the CIAO spectral fitting tool {\tt Sherpa} to analyze the resulting
data.  Background levels are negligible ($\leq 1$ count expected in the
source extraction region) in all but two observations; for these two, we
fit the background simultaneously with a power law model.
For the twelve objects with sufficient counts to obtain binned
spectra (with at least 5 counts per bin), we fit a simple power law
with foreground Galactic absorption taken into account\footnote{
For fitting we used the  $\chi^2$ statistic with the \citet{gehrels86} 
variance function, which is the {\tt Sherpa} default and applicable
to data with few counts.}.  This model can be represented by the equation
\begin{equation}
dN(E) \propto E^{-\Gamma} e^{-N_H \sigma(E)} A(E) dE
\end{equation}
where $\sigma(E)$ is the photoelectric absorption cross section of
Galactic gas with effective column density $N_H$, and $A(E)$ is the
observation--specific effective detector area encoded within each
ancillary response function (ARF) file used within {\tt Sherpa}.
Knowing $N_H$, $\Gamma$ and the overall normalization can be found
through a simple two--parameter fit to the data.

For those objects too faint to be analyzed as binned spectra in {\tt
Sherpa} (and to check the consistency of our spectral fits) we employed
a hardness ratio fitting method.  This method reduces the problem to a
one--parameter fit by disregarding the overall normalization of the
spectrum (which can be determined afterwards from the total count rate,
$(H+S)/t_{\rm exp}$).
The hardness ratio is defined as ${\rm HR} = (H-S)/(H+S)$,
where $H$ and $S$ are now the net counts in some arbitrary hard and soft
bands respectively.  Assuming Poisson errors, $\sigma_{\rm
HR}=2\sqrt{S^2 H+H^2 S}/(H+S)^2$ (again using the {\it raw} number of
counts per band to calculate errors, for those observations with
non--zero background levels).  We employed two different hardness ratios
in this analysis, which we call HR$_a$ and HR$_b$.  HR$_a$ is analogous
to the ROSAT ``hardness ratio 2'' parameter, where the soft and hard
bands are 0.5--0.9 and 0.9--2~keV respectively.  HR$_b$ takes advantage
of the full \chandra\ energy range with soft and hard bands covering
0.5--2 and 2--8~keV, respectively.  Again using the ARF and $N_H$ value
specific to each observation, we used {\tt Sherpa} to calculate HR$_a$
and HR$_b$ for a test value of $\Gamma$.  By iterating $\Gamma$ until
the correct values of the hardness ratios were reached, we derived
photon indices independently for each of HR$_a$ and HR$_b$.

\subsection{Consistency}
For 15 of these NLS1s, $\Gamma$ was derived by using two or three of the
methods discussed in the previous section.  As a consistency check, we
compared the results for individual objects.  For the most part,
especially for the brighter objects, the three measurements produced
results well within $1\sigma$ of each other.  However, there are
systematic offsets.  Values of $\Gamma$ derived from HR$_a$ are nearly
always higher than those found using HR$_b$, while {\tt Sherpa} fits
typically lie somewhere in between. This is as expected since the soft
X-ray power-law slopes of NLS1s are known to be steeper than those in
the hard band due to the commonly--observed soft excess 
(e.g. ROSAT: BBF96 vs. ASCA: Brandt, Mathur \& Elvis 1997, although some
of this offset may be due to calibration issues as described in Iwasawa, 
Fabian, \& Nandra 1999).
A slope over the entire energy range, as measured
by {\tt Sherpa}, would be in between the two. Since they contain the
average spectral information, we list the $\Gamma$ from {\tt Sherpa}
fits in Table~\ref{tab_objects}.  If there are not enough counts for
such a fit, we list the 0.5--8~keV (HR$_b$) measurement since it covers
the full energy range; for the lone observation with no 2--8~keV counts,
the 0.5--2~keV measurement of $\Gamma$ is listed.


\section{X-ray Spectral Properties}

\subsection{Detectability in the RASS\label{sec_rass}}
The NLS1s in this sample span a broad range of $\Gamma$ (best fit values
of 0.25--3.15) and X-ray fluxes from near the detection limit of about
0.002 counts~s$^{-1}$ to almost 0.2 counts~s$^{-1}$.  Taking the
0.5--2~keV flux derived from {\tt Sherpa} for the twelve brightest
objects, we used the PIMMS software\footnote{Portable, Interactive
Multi-Mission Simulator Version 3.4, from NASA's High Energy
Astrophysics Science Archive Research Center, currently available at
\url{http://heasarc.gsfc.nasa.gov/docs/software/tools/pimms.html}.} to 
calculate the count rate expected in ROSAT.  
Exposure times for each position were taken from the RASS exposure
maps\footnote{Available at 
\url{http://www.xray.mpe.mpg.de/cgi-bin/rosat/rosat-survey}.} and the 
expected number of RASS counts was then computed for each object.  Three
of the sources, J1013$+$0102, J1214$+$0055, and J1449$+$0022 should have
been easily detected with about 17, 17, and 22 counts, respectively.
The second object may actually correspond to a nearby RASS source which
was marginally detected and not cross-referenced in the SDSS database.
Four other objects in this sample should be marginally detected by the
RASS with $\sim 10$ counts; their nondetection may be due to Poisson
error or uncertainty in the flux determination.  The remaining ten
objects were well below the RASS detection limit.

The X-ray luminosity of the three bright objects may have varied by a
factor of two or more between their RASS and \chandra\ observations,
which could account for their lack of detection in the RASS.  Such
long--timescale variability is not surprising, as it is commonly seen in NLS1s
\citep[e.g.,][]{grupe01}.  It is impossible to ascertain the
degree of variability in the fainter objects; however, most of this
sample could not have been detected in the RASS at the flux levels
observed by \chandra.  It is unlikely that all 14 of these 17 NLS1s were
in an exceptionally low luminosity state during both the RASS and
\chandra\ observations; thus, at least some of these objects must be
intrinsically faint at ROSAT energies.

\subsection{Photon Index--Luminosity Relation}

It is likely that some combination of variability, X-ray hardness, and
low X-ray luminosity resulted in these objects not being detected in the RASS.
The latter two factors can fortunately be determined from these 
observations.  The data in Table~\ref{tab_objects} show that many of 
the NLS1s with low $\Gamma$ are also the faintest X-ray sources.
To determine whether this is a true effect, we calculated the monochromatic
1~keV (rest-frame) fluxes using the models derived through spectral
fitting in {\tt Sherpa}.  For the fainter objects we calculated the 
expected flux from the best--fit hardness ratio models using {\tt Sherpa}, 
with a fractional
flux error equal to $N^{-1/2}$ where $N$ is the total number of counts
in the 0.5--8~keV band.
Using the SDSS redshifts and an $\Omega_m=0.3$, $\Omega_\Lambda=0.7$, 
$H_0=70\,\rm{km}\,\rm{s}^{-1} \rm{Mpc}^{-1}$ cosmology, the corresponding
rest--frame luminosities were then calculated.  These are given in 
Table~\ref{tab_objects}, along with the optical $\lambda 5100$\AA\ 
monochromatic flux measured from the original SDSS spectra.

\begin{figure}
\plotone{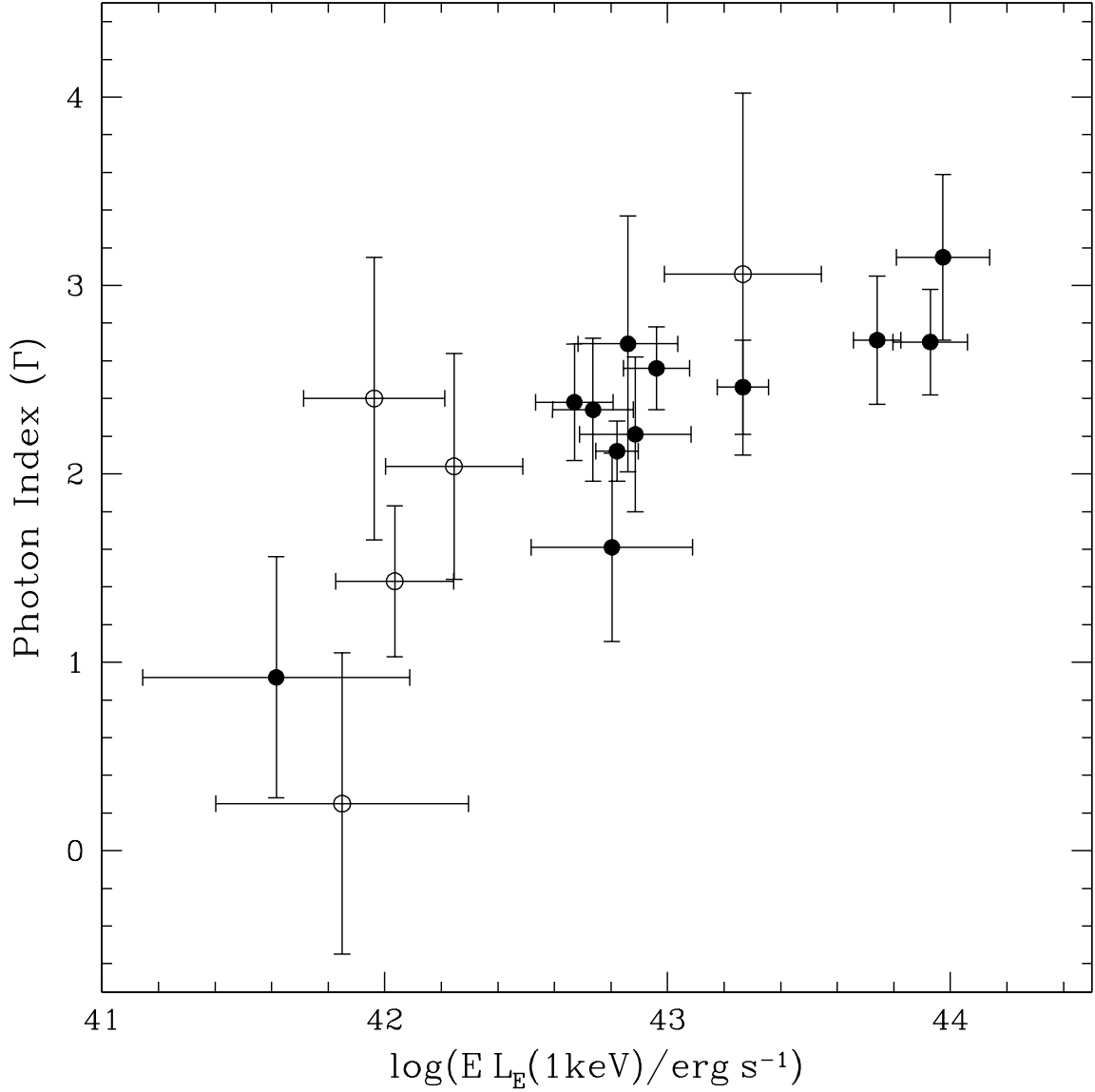}
\caption{$\Gamma$ (from Table~\ref{tab_objects}) vs. 1~keV rest--frame
luminosity for the \chandra\ follow--up NLS1 sample.  Filled points 
are bright objects with spectra fit in Sherpa, while for open circles 
the luminosity was estimated using HR$_b$ (HR$_a$ for J1311$+$0031) and 
the 0.5--8~keV count rate. \label{fig_gamlum}}
\end{figure}

The relation between $\Gamma$ and $\l1kev$ is shown in Figure~\ref{fig_gamlum}.
Luminosities derived from {\tt Sherpa} fitting are plotted as filled
circles, and those estimated from the hardness ratio models are plotted
as open circles.  This figure shows a strong correlation between X-ray
spectral slope and 1~keV luminosity.  \citet{grupe99} found such a
correlation among soft X-ray selected AGN, with a stronger correlation
among NLS1s noted in \citet{grupe04-2}.  No such correlation is seen
in their broad--line AGN sample; additionally, \citet{laor97} do not find
a significant correlation in their sample of bright, optically--selected 
quasars (although this may be due to the limited range of $\Gamma$ in these
samples).

This relation has a slope of $b=0.66\pm 0.13$ (where $\Gamma
\propto b \log \l1kev$); if the two hardest objects are excluded
from the fit, then $b$ decreases to $0.55\pm 0.13$.  Similar relations
have been measured for NLS1s by \citet{forster96} with $b=0.32$ 
(where the luminosity
is measured over the 0.1--2~keV range), while the \citet{dai04} quasar sample
has $b=0.55\pm 0.11$ (with $L=L_{0.2-2\,\rm{keV}}$).  When the luminosities
of our sample are extrapolated to the 0.1--2\,keV range using the best--fit
models and the two lowest--$\Gamma$ points are excluded, the slope
becomes $b=0.48\pm 0.08$, in agreement with \citet{dai04} but about
2$\sigma$ higher than the \citet{forster96} result (note that the choice of
0.1\,keV or 0.2\,keV as the lower bound of the energy range
does not significantly affect the extrapolated luminosity).
This latter discrepancy may be due to their sample being selected from the
RASS, which is not as sensitive to flat--spectrum objects as \chandra.

For contrast, the $\Gamma-\l1kev$ relation caused by pure
absorption (i.e. assuming a typical source with $\Gamma\sim 2.5$, determining
what hardness ratio and $\l1kev$ would be observed with
various intrinsic $N_H$ values, and recalculating $\Gamma$ using the simulated
HR$_b$ assuming \emph{no} prior knowledge of the intrinsic absorption)
produces a slope closer to $b=2$ for $N_H\la 5\times 10^{21}$\,cm$^{-2}$.  
Most of the observed correlation thus
cannot be explained simply by intrinsic absorption, although it could
explain the hardest objects in this sample ($\Gamma < 1$).  However, this
would require large neutral column densities ($\ga 10^{22} \rm{cm}^{-2}$), 
despite there being no sign of optical dust reddening.  Moreover, it is
unlikely that all four NLS1s with $\Gamma < 1$ are strongly absorbed
(see \S\ref{sec_abs}).  A dust--poor absorber could, in
principle, account for the extreme hardness and low X-ray luminosity of 
one or two of these objects.

\begin{figure}
\plotone{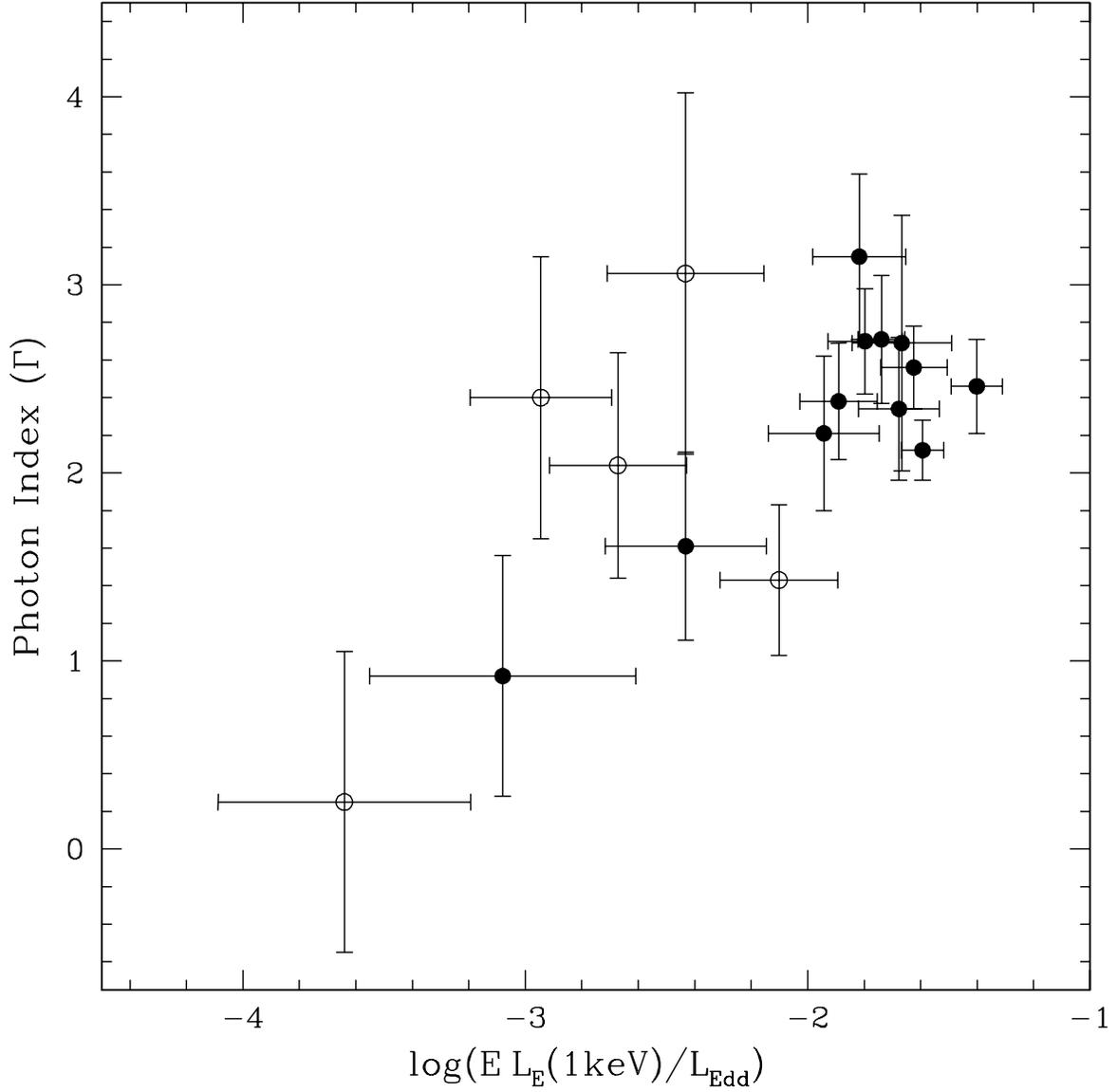}
\caption{Same as Figure~\ref{fig_gamlum}, but with photon index plotted
against $\l1kev/L_{\rm Edd}$.  The x-axis errorbars only reflect
the uncertainty on $\l1kev$ since the uncertainty in 
$L_{\rm Edd}$ is not well-known. \label{fig_gamledd}}
\end{figure}

\begin{figure}
\plotone{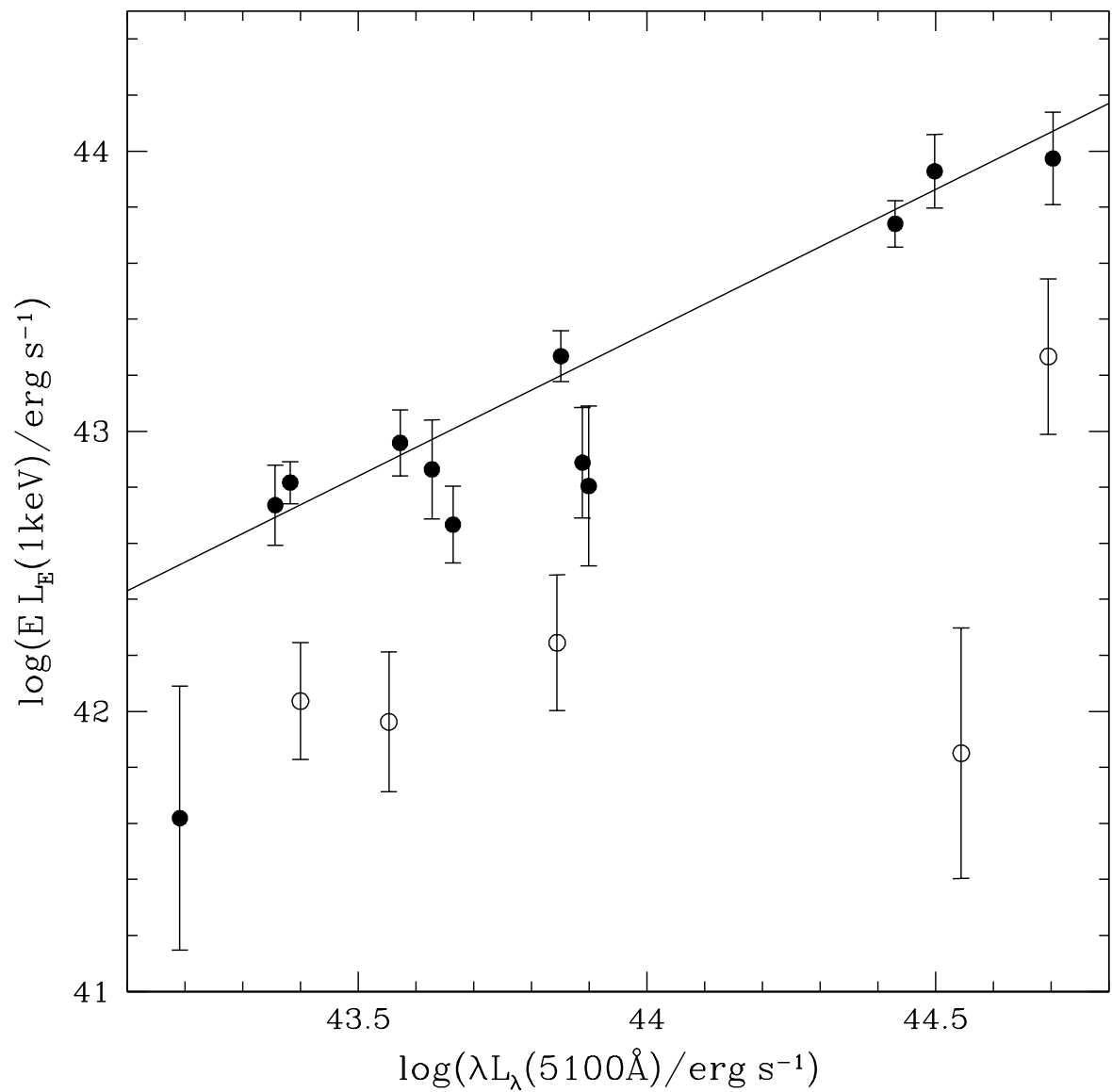}
\caption{X-ray vs. optical luminosity for the objects with $\Gamma$ derived
in {\tt Sherpa} (solid points), or using hardness ratios (open circles).
The solid line denotes the best--fit $\l1kev-L_{\rm opt}$ relation
from the WPM02 RASS sample.
The high--optical, low--X-ray luminosity point is the object with 
$\Gamma=0.25$, and may be highly absorbed. \label{fig_optx}}
\end{figure}

The observed $\Gamma-\l1kev$ correlation may be a result of
differing Eddington ratios.  To test this, we estimated black hole masses
and Eddington luminosities using the \citet{kaspi00} 
relation between $R_{\rm BLR}$, $\lambda L_\lambda$(5100\AA), and
FWHM(H$\beta$) and applying the virial theorem.  It should be noted that
these estimates may be subject to systematic error since the \citet{kaspi00}
sample contains very few NLS1s; however, this method should be sufficient
to show general trends in the data.  Values of FWHM(H$\beta$)
were taken directly from WPM02, and the 5100\AA\ flux was measured from
the SDSS spectra used in that study.  Figure~\ref{fig_gamledd} demonstrates
a strong correlation between $\Gamma$ and $\l1kev/L_{\rm Edd}$, which
further indicates that the objects with the softest X-ray spectra
have the highest relative accretion rates if $\l1kev\propto L_{\rm bol}$.
This is probably a reasonable assumption.  Figure~\ref{fig_optx} shows
that the 1~keV luminosity is correlated with the 5100\AA\ luminosity
for the bright and faint objects in this sample (with the notable exception
of one point, which has $\Gamma\sim 0.25$ and may be strongly absorbed).
We can thus assume that both the optical and X-ray luminosities
provide some indication of the bolometric luminosity.  As expected, a
correlation is also seen between $\Gamma$ and $\lambda L_\lambda$(5100\AA),
though it is somewhat weaker than that between $\Gamma$ and $\l1kev$.
This weakness is not surprising, since X-ray properties should physically be
more closely linked with each other than with optical measurements.
 
Although $\Gamma$ is directly proportional to $L_{\rm bol}/L_{\rm Edd}$, 
it is thought to be inversely proportional to $\mbh$ \citep[e.g.,][]{kuras00}, 
which could wash out the observed $\Gamma-\l1kev$
correlation for some samples.  For this \chandra\ sample, the $\mbh$ 
estimates span approximately two orders of magnitude, while 
$\l1kev/L_{\rm Edd}$ spans three, so the correlation is 
observed.  In samples with a larger range of black hole masses and luminosities
(such as the Grupe 2004 BLS1s), this correlation would not be expected
and, indeed, is not seen.
The $\Gamma-\l1kev$ relation would also not be seen in samples with
a smaller range of $L_{\rm bol}/L_{\rm Edd}$, unless the black hole mass
range is correspondingly larger, as proposed for the \citet{dai04} sample.  
When the same analysis is performed
for the WPM02 RASS objects, no correlation is observed.  Instead, they cluster
almost exclusively around the highest--$\l1kev/L_{\rm Edd}$, 
highest--$\Gamma$ members of the \chandra\ sample.  This is not unexpected
since the WPM02 RASS sample consists mostly of NLS1s near the RASS 
detection limit; they do not exhibit the range in luminosities
(and hence $\l1kev/L_{\rm Edd}$) or $\mbh$ required to see the 
$\Gamma-\l1kev$ correlation.  If the observed $\Gamma-\l1kev$ correlation is 
indeed due to differences in the Eddington ratio, then it is 
quite likely that this \chandra\ sample includes some NLS1s which are 
accreting far below the Eddington limit.

\subsection{Spectral Features in J1449+0022\label{sec_771}}
\begin{figure}
\plotone{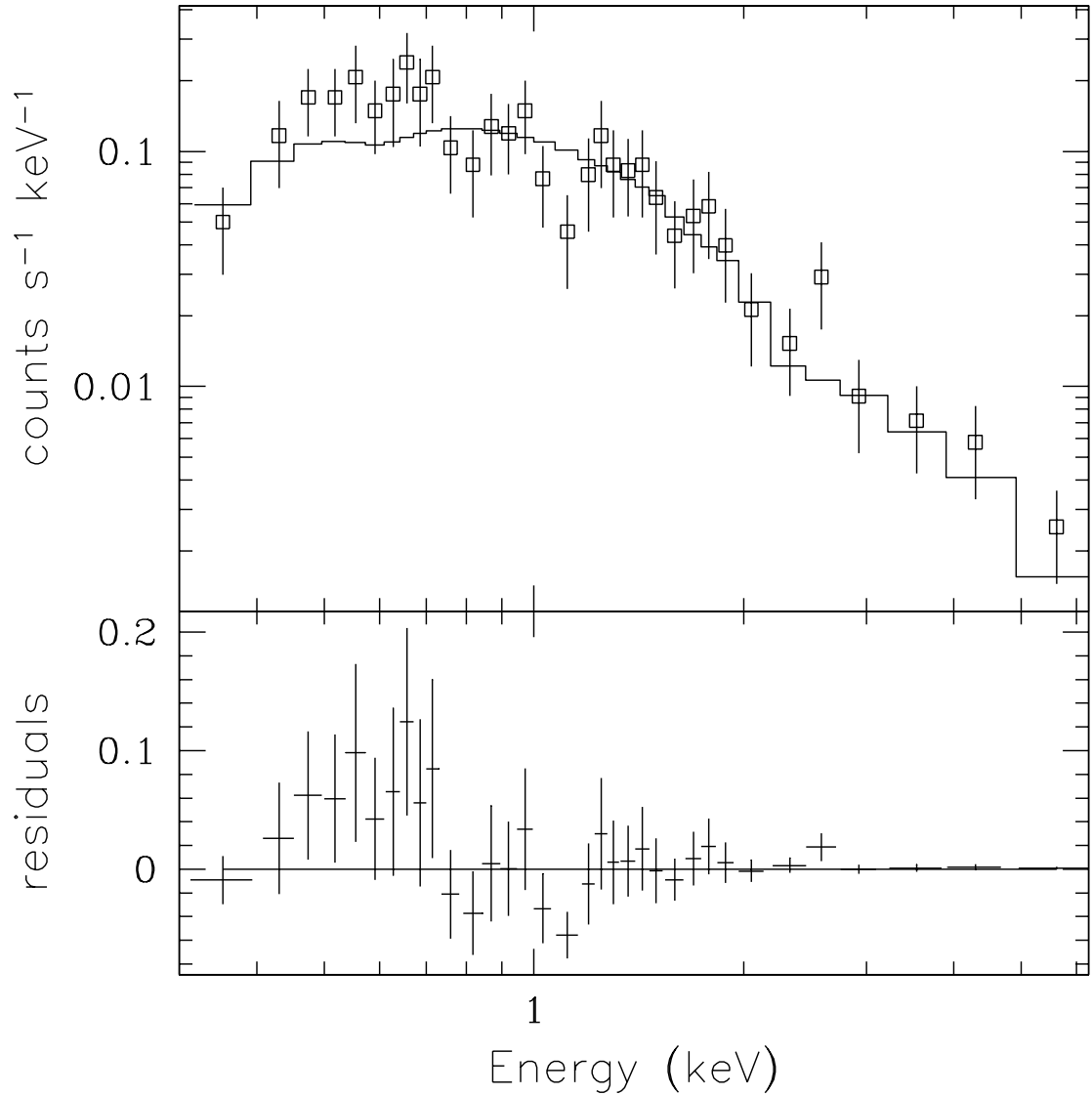}
\caption{\chandra\ count spectrum for J1449+0022, along with the best-fit
power-law model and residuals.  This spectrum has been binned to 10
counts per bin.  The observed flux is systematically high at $E<0.8\rm{keV}$,
and there is an absorption feature just above 1\,keV. \label{fig_771}}
\end{figure}

As noted in \S\ref{sec_rass}, J1449+0022 is the brightest object (in
X-rays) in this sample and may have gone undetected in the RASS due to
strong variability.  Indeed, there appears to be a slight decrease in
the \chandra\ count rate over the course of the 2~ksec observation.
More interesting, however, is that the spectrum of this object is not
well fit by a simple power law with Galactic foreground absorption, as
seen in Figure~\ref{fig_771}; specifically, there appears to be an
excess of soft photons over that expected from a power law.  The X-ray
continuum can be better fit using a slightly more complicated model;
either a power law plus a blackbody (which yields $\Gamma=1.58\pm 0.22$
and $kT=0.11\pm 0.02$), or a broken power law with
$\Gamma_1=3.40^{+0.73}_{-0.44}$, $\Gamma_2=1.64\pm 0.19$, and
$E_b=1.04^{+0.12}_{-0.19}$\,keV.  There is also a significant dip in
flux at $1.10\pm 0.03$\,keV with equivalent width $0.12^{+0.06}_{-0.04}$\,keV. 
This absorption is similar to that seen in some other NLS1s
\citep{leighly97}.  Such a feature could be due OVII/OVIII
absorption in a highly relativistic outflow \citep[e.g.,][]{pounds03}, 
but a more plausible model
may be the \citet{nicastro99} hypothesis of a strong Fe~L complex at 
this energy.

\subsection{Intrinsic Absorption \label{sec_abs}}
Four of the NLS1s in this sample exhibit spectra with $\Gamma < 2$.
Since for low count rates $\Gamma$ and $N_H$ are highly degenerate, no
individual object is bright enough to place a meaningful limit on its
intrinsic absorption; thus, we cannot determine with confidence whether
this X-ray hardness is intrinsic to the accretion process or merely
caused by a high degree of absorption.  The spectra of these four NLS1s
were added together to determine whether they exhibit strong absorption
as a group.  We assumed all have typical $\Gamma =2.5$ and similar
$N_H$ values at each redshift, and fit a corresponding model to the coadded
spectrum.  This resulted in a best-fit value of
$N_H=1.5^{+1.6}_{-0.9}\times 10^{21} \rm{cm}^{-2}$, but the fit is not
particularly good and shows strong residuals.  If we leave both $\Gamma$
and $N_H$ as free parameters, $N_H$ becomes zero (with a $2\sigma$ upper
limit of $2\times 10^{21} \rm{cm}^{-2}$) and $\Gamma = 1.2\pm 0.3$.  The
fit with no absorption is also much better ($\Delta \chi^2 = -6.6$
compared to the absorption model).  Based on this, it appears as though
these flat--spectrum sources are not strongly absorbed as a group.

Of course, it is possible that one or two of these NLS1s could be
strongly absorbed, particularly the faintest object (J1259$+$0102) since
with only five total counts its influence on the coadded spectrum is
small.  If this object has an intrinsic X-ray luminosity 
$\l1kev\sim 10^{44}\,\rm{erg}\,\rm{s}^{-1}$ and $\Gamma\sim 3$ 
(similar to the brightest objects in our sample), an
intrinsic neutral column density of $\sim 2.5\times 10^{22}
\rm{cm}^{-2}$ would be required to reproduce the observed flux and
$\Gamma$.  The input $\Gamma$ and luminosity of such a model can be
adjusted to reproduce the other faint, hard sources as well, but
somewhat high column densities ($N_H \ga 5\times 10^{21} \rm{cm}^{-2}$)
are typically required.  Moreover, there is no indication of absorption
in the optical spectra of these objects.  Thus, we conclude that these
objects most likely have intrinsically hard X-ray spectra, although the
two hardest may be heavily obscured in the X-rays but not at optical
wavelengths \citep[cf.][]{risaliti01,nandra04}.

\section{Comparison to Other NLS1 Samples}
\subsection{X-ray Spectral Slope}
At first glance, this sample appears to exhibit a range of X-ray
spectral slopes extending to much lower values than previously seen for
NLS1s.  However, the $\Gamma$ values listed are, for the most part,
derived over a larger energy range than used by previous studies.  For
example, $\Gamma$ is measured by ROSAT over the 0.1--2.4~keV energy
range for the 52 narrow- and broad-line Seyfert~1 objects in BBF96,
while the 0.1--2.0~keV hardness ratio is used with the WPM02 sample.  In
order to ensure the energy bands overlap, we employ the HR$_a$
measurement of $\Gamma$ for comparison to these two samples.

It should be noted that the energy ranges used by these samples are
slightly different nonetheless.  Since HR$_a$ covers the 0.5--2.0~keV
band, it does not extend to the 0.1~keV minimum energy of ROSAT, or to
the 2.4~keV maximum energy used by BBF96.  However, 28 of the WPM02
NLS1s detected in ROSAT have measurements for both HR1 (0.1--2.0~keV)
and HR2 (0.5--2.0~keV).  We rederived $\Gamma$ using the HR2
measurements and compared it to the values from HR1.  Although the
errors are larger, the HR2 $\Gamma$s are consistent with those from HR1
in all but four objects.  No systematic offset is seen between these two
measurements.  Thus, we assume that the $\Gamma$s derived from ROSAT HR1
are good estimators of the spectral slopes over the 0.5--2~keV range.
As for the BBF96 measurements extending to 2.4~keV while ours extend
only to 2.0~keV, we expect this to be negligible for two reasons: first,
few photons are emitted in the high--energy tail ($N_{2.0-2.4{\rm keV}}
\sim 0.1N_{0.5-0.9\rm{keV}}$ for $\Gamma=2$); and second, the ROSAT
sensitivity decreases quite rapidly in this regime, so even fewer
photons should be detected.  Thus, the differing energy bands should
have little effect on comparing these various samples.

Figure~\ref{fig_fgamma} shows the relation between $\Gamma$ and the
H$\beta$ velocity width for all three samples.  Crosses denote
BBF96 data, blue circles are the WPM02 RASS data, and the 16 \chandra\
objects with HR$_a$ measurements are shown as red circles with errorbars.  
The horizontal lines show the mean and sample standard deviation of 
$\Gamma$ for only the BBF96 data; NLS1 and Sy1 averages are computed 
separately.  It is immediately apparent that the WPM02 and \chandra\ 
NLS1s do not fall within the same range as those in BBF96; in fact, 
a significant number appear to exhibit \emph{Seyfert 1}--type
X-ray spectra.  Table~\ref{tab_gamma} lists the mean and sample standard
deviation of $\Gamma$ for the three samples.  An interesting progression
emerges: as we move from NLS1s which were found through a mixture of
optical and X-ray selection (BBF96), to those which were selected solely 
based on their optical properties (WPM02), to the \chandra\ sample
presented herein which was selected on the basis of weak X-ray emission,
$\langle\Gamma\rangle$ and the extremes of the $\Gamma$ distribution become
harder.  In fact, the $\Gamma$ distribution of this sample extends to
lower values than that of the BBF96 Seyfert~1 galaxies.  This illustrates
the effect of selection methods on building NLS1 samples, as well as
demonstrating the existence of a significant population of NLS1s which 
are {\emph not} ultrasoft X-ray sources.

\subsection{X-ray vs. Optical Luminosity}
This \chandra\ sample was originally selected as a set of X-ray weak
NLS1 candidates, i.e. NLS1s which were not detected in the RASS but 
had $g^\prime$ magnitudes similar to detected objects.  To determine
whether these objects are truly X-ray weak, we used the WPM02 RASS--detected
NLS1s as a comparison sample.  Rest--frame luminosities at 
1\,keV were calculated using PIMMS along with ROSAT count rates and 
the best--fit $\Gamma$ values given in WPM02, and 5100\,\AA\ luminosities
were measured from the original SDSS spectra.  The best--fit relation 
between the monochromatic optical and X-ray luminosities has a slope
consistent with unity, and is plotted on Figure~\ref{fig_optx} along
with the data points from the \chandra\ sample.

Interestingly, the RASS relation corresponds very well with the
upper bound on the X-ray--optical relation for the \chandra\ NLS1s.
Eight of these objects are consistent with the RASS fit and another 
three are consistent within $\sim 1-2\sigma$, while the
rest are significantly below the line.  The faintest objects in X-rays
appear to exhibit a $L_X-L_{\rm opt}$ slope similar to that of the
brighter objects (if the point at the lower right of the plot, which
is probably absorbed in X-rays, is excluded), but their X-ray luminosities
are roughly a factor of 5 lower than those seen in the WPM02 RASS sample.
Thus, although most of the objects in this sample are not particularly
X-ray faint compared to NLS1s in the WPM02 RASS sample with similar optical
luminosities, six of them do fall well below the $L_X-L_{\rm opt}$
relation due to intrinsic X-ray faintness or obscuration.

\section{Discussion and Conclusions\label{conclusions}}

Through short--duration \chandra\ observations of RASS--undetected NLS1s, 
we have determined that six of the 17 objects have X-ray luminosities
substantially lower than NLS1s with similar optical properties.  Of
the brighter objects, at least two exhibit flux levels which should
have been detectable by the RASS, indicating that their luminosities may
have increased by a factor of two or more between the RASS and \chandra\
observations.  Many of the remaining bright objects were near or just
below the RASS detection limit, and were most likely not seen due to
smaller luminosity variations or Poisson noise.  
Across the entire sample, a strong correlation is seen 
between the X-ray spectral slope $\Gamma$ and $\l1kev$.  This is probably
not entirely due to intrinsic absorption, since individual spectra of 
bright objects
as well as a coadded spectrum of the faintest objects do not indicate
high degrees of absorption (although one or two of the faintest hard--spectrum
objects may be absorbed in X-rays but not at optical wavelengths).  
If $\Gamma$ is indeed correlated with $L_{\rm bol}/L_{\rm Edd}$, then
the $\Gamma-\l1kev$ relation suggests that variations in $\l1kev$ are 
primarly due to
differences in $L_{\rm bol}/L_{\rm Edd}$ among objects with comparatively
similar black hole masses.  This interpretation is complementary to that
of \citet{dai04}, who find a similar relation but whose sample more 
likely includes objects with a large
range of $\mbh$ but $L_{\rm bol}/L_{\rm Edd}\sim 1$.  

These observations may hold important implications for the ``Eigenvector 
1'' (Principal Component 1; PC1) paradigm
posited by \citet{bg92} and reinforced by \citet{boroson02}.  In this
picture, PC1 (which is primarily driven by an anticorrelation between
[\ion{O}{3}] and \ion{Fe}{2}) is an indicator of $L_{\rm bol}/L_{\rm Edd}$.  
NLS1s typically lie at one extreme end of PC1---the 
end thought to correspond to the highest relative accretion rates.  Since
$\Gamma$ is thought to be related to $L_{\rm bol}/L_{\rm Edd}$, $\Gamma$ and
PC1 should be correlated; indeed, \citet{brandt98} find such a correlation.
However, the sample presented herein contains objects which from their 
optical spectra are at the
supposed high--$L_{\rm bol}/L_{\rm Edd}$ end of PC1, yet also exhibit very low
values of $\Gamma$, as well as low inferred $\l1kev/L_{\rm Edd}$.  

These extreme objects may indicate that while PC1 is usually
correlated with $L_{\rm bol}/L_{\rm Edd}$, it may also be 
affected by orientation, black hole mass, or other physical drivers 
\citep[as noted by][]{boroson04}.  This is not
a completely new phenomenon; for example, BBF96 note that Mrk~507, with
$\Gamma=1.6\pm 0.3$, has an unusually flat X-ray spectrum for a NLS1.  
This sample simply demonstrates
that Mrk~507 is not an isolated case, and in fact a small but interesting
subset of NLS1s do not appear to fit within the PC1 framework.  The
apparent lack of strong absorption in some of these flat sources indicates
that their X-ray spectra actually are intrinsically flat.  
Further studies of X-ray weak NLS1s, as well as much larger samples from
surveys such as the SDSS, should offer greater insight into the 
mechanism(s) behind PC1.

Due to the short exposure times of these observations ($\la 2$~ksec), we 
cannot infer much outside
of $\Gamma$ and luminosity estimates for individual objects; indeed, this
program was intended to study the group properties of an X-ray weak
NLS1 sample.  However, there are several objects in this sample with
exceptionally low $\Gamma$ which may be worthy of further study.
These hard X-ray NLS1s may represent a new, rare
subclass which are optically normal but highly absorbed in the X-rays,
or which exhibit abnormally low $L_{\rm bol}/L_{\rm Edd}$, or both.

\begin{figure}
\epsscale{0.8}
\plotone{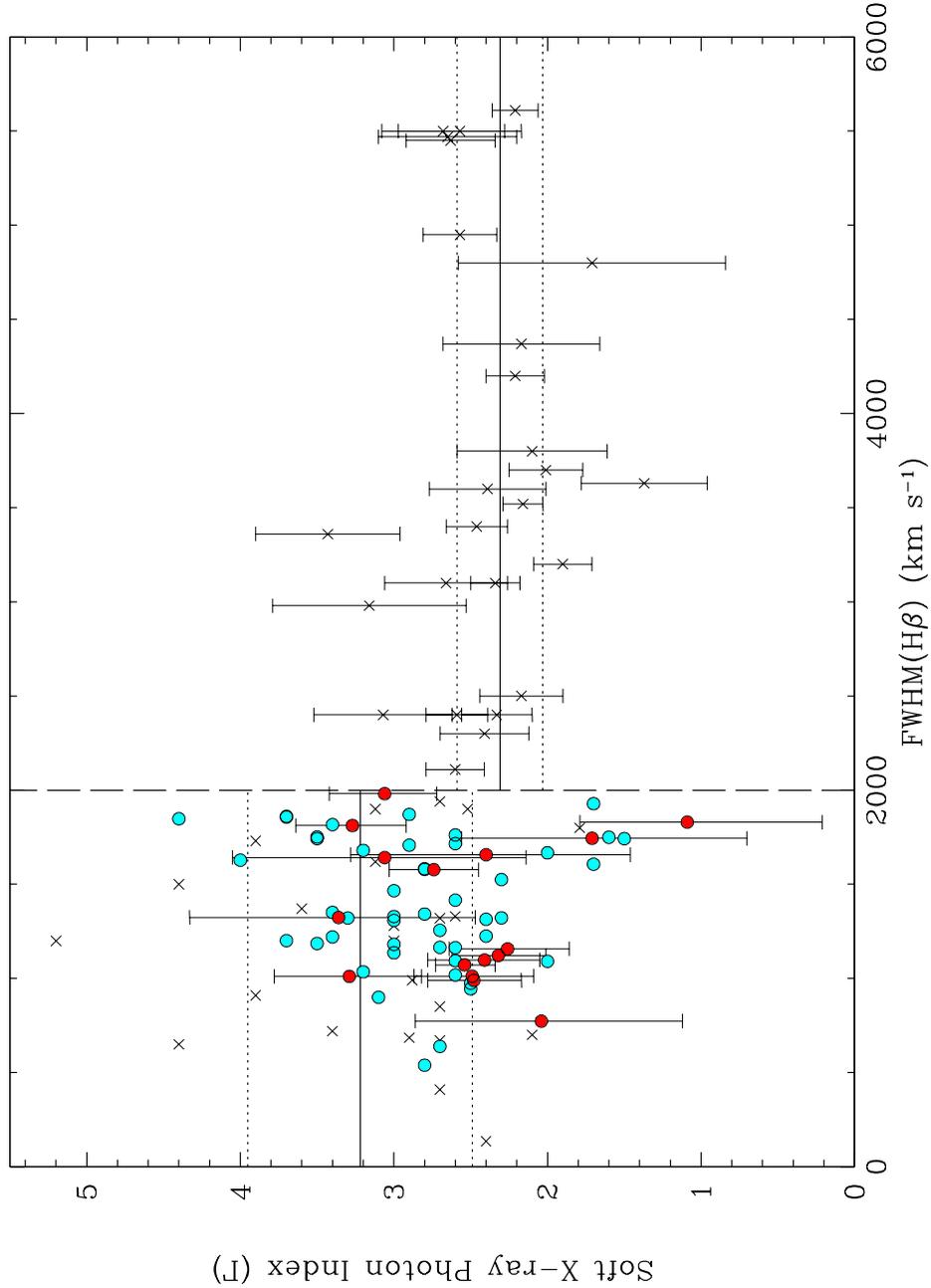}
\caption{Soft X-ray photon index vs. H$\beta$ width, for NLS1s and Seyfert~1
galaxies (on the left and right of the dashed vertical line, respectively). 
Crosses are data from BBF96, kindly provided by Th. Boller.  Blue circles
are NLS1s from WPM02 (with $\Gamma$ derived from RASS measurements) and 
red circles are the sample presented herein.  All values of $\Gamma$ for the
\chandra\ follow--up sample were found using the 0.5--2~keV hardness ratio
HR$_a$.  The
horizontal solid and dashed lines show the means and sample standard deviations
of $\Gamma$ for the BBF96 NLS1s and Sy1s. \label{fig_fgamma}}
\end{figure}

\acknowledgements
The authors are grateful to Th. Boller for providing the original BBF96 data 
for our Figure~\ref{fig_fgamma}.  We also thank the anonymous referee
for helpful comments and suggestions.  
Support for this work was provided by NASA through Chandra Award Number
GO3-4145X issued by the Chandra X-ray Observatory Center, which is
operated by SAO for NASA under contract NAS8-39073.

\clearpage

\clearpage

\begin{deluxetable}{lccccccc}
\tablecolumns{8}
\tablewidth{470pt}
\tablecaption{\chandra\ observing log\label{tab_obslog}}
\tablehead{
\colhead{SDSS Name\tablenotemark{a}} &
\colhead{Date} &
\colhead{$t_{\rm exp}$} &
\colhead{S\tablenotemark{b}} &
\colhead{M\tablenotemark{b}} &
\colhead{H\tablenotemark{b}} &
\colhead{$N_H$\tablenotemark{c}} &
\colhead{CR\tablenotemark{d}}
\\
\colhead{} &
\colhead{} &
\colhead{(s)} &
\colhead{} &
\colhead{} &
\colhead{} &
\colhead{($10^{20}$\,cm$^{-2}$)} &
\colhead{($s^{-1}$)}
}

\startdata
J002305.03$-$010743.5 &25--08--2003 &1940 &30 &44 &13 &2.81 &$0.045\pm 0.005$\\
J002752.39$+$002615.8 &03--09--2003 &1560 &7  &21 &14 &2.72 &$0.027\pm 0.004$\\
J015652.43$-$001222.0 &12--09--2003 &1710 &8  &6  &3 &2.62  &$0.010\pm 0.002$\\
J022756.28$+$005733.1 &23--06--2003 &1730 &6  &10 &7 &2.84  &$0.013\pm 0.003$\\
J022841.48$+$005208.6 &24--06--2003 &2000 &51 &65 &22 &2.76 &$0.069\pm 0.006$\\
J031427.47$-$011152.4 &03--09--2003 &1910 &48 &43 &9 &5.70  &$0.052\pm 0.005$\\
J101314.86$-$005233.5 &08--01--2003 &1990 &63 &70 &14 &3.64 &$0.074\pm 0.006$\\
J104230.14$+$010223.7 &20--02--2003 &1730 &33 &42 &12 &3.72 &$0.050\pm 0.005$\\
J121415.17$+$005511.4 &07--02--2003 &1960 &49 &42 &12 &1.94 &$0.053\pm 0.005$\\
J125943.59$+$010255.1 &04--03--2003 &1940 &0  &2  &3 &1.62  &$0.003\pm 0.001$\\
J131108.48$+$003151.8 &10--03--2003 &1730 &7  &6  &0 &1.90  &$0.008\pm 0.002$\\
J141234.68$-$003500.0 &07--01--2003 &2090 &36 &48 &16 &3.29 &$0.048\pm 0.005$\\
J143030.22$-$001115.1 &23--04--2003 &1950 &5  &10 &9 &3.15  &$0.012\pm 0.003$\\
J144932.70$+$002236.3 &09--07--2003 &2150 &129 &165 &68 &3.75 &$0.168\pm 0.009$\\
J145123.02$-$000625.9 &22--04--2003 &2120 &52 &74 &14 &3.84 &$0.066\pm 0.006$\\
J170546.91$+$631059.1 &17--09--2003 &1940 &6  &8  &2 &2.57  &$0.008\pm 0.002$\\
J233853.83$+$004812.4 &28--08--2003 &1910 &28 &23 &11 &3.88 &$0.032\pm 0.004$\\

\enddata

\tablenotetext{a}{Format: SDSS JHHMMSS.ss$\pm$DDMMSS.s}
\tablenotetext{b}{Net counts in soft (0.5--0.9~keV), medium
(0.9--2.0~keV), hard (2.0--8.0~keV) bands.}
\tablenotetext{c}{Galactic foreground \ion{H}{1} column density,
found with the {\tt nh} utility.}
\tablenotetext{d}{Net \chandra\ 0.5--8\,keV count rate.}

\end{deluxetable}

\begin{deluxetable}{lccccccc}
\tablecolumns{8}
\tablewidth{500pt}
\tablecaption{NLS1 Optical and X-ray Properties\label{tab_objects}}
\tablehead{
\colhead{Name\tablenotemark{a}} &
\colhead{z\tablenotemark{b}} &
\colhead{FWHM(H$\beta$)\tablenotemark{c}} &
\colhead{$\Gamma$\tablenotemark{d}} &
\colhead{$\log(E L_E)$\tablenotemark{e,g}} &
\colhead{$\log(\lambda L_\lambda)$\tablenotemark{f,g}} &
\colhead{$\log(\frac{\mbh}{M_\sun})$\tablenotemark{h}} &
\colhead{Note}
\\
\colhead{} &
\colhead{} &
\colhead{(km\,s$^{-1}$)} &
\colhead{} &
\colhead{(erg\,s$^{-1}$)} &
\colhead{(erg\,s$^{-1}$)} &
\colhead{} &
\colhead{}
}

\startdata
J0023$-$0107 &0.166 &1160 &$2.21\pm 0.41$         &$42.89\pm 0.20$ &$43.89$ &6.73 &\nodata\\
J0027$+$0026 &0.205 &1830 &$1.61\pm 0.50$         &$42.80\pm 0.29$ &$43.90$ &7.14 &\nodata\\
J0156$-$0012 &0.163 &1320 &$2.04^{+0.75}_{-0.49}$ &$42.24\pm 0.24$ &$43.84$ &6.82 &1\\
J0227$+$0057 &0.128 &770  &$1.43\pm 0.44$         &$42.04\pm 0.21$ &$43.40$ &6.04 &1\\
J0228$+$0052 &0.186 &990  &$2.46\pm 0.25$         &$43.27\pm 0.09$ &$43.85$ &6.57 &\nodata\\
J0314$-$0111 &0.387 &1810 &$3.15\pm 0.44$         &$43.97\pm 0.16$ &$44.70$ &7.69 &\nodata\\
J1013$-$0052 &0.276 &1580 &$2.71\pm 0.34$         &$43.74\pm 0.08$ &$44.43$ &7.38 &\nodata\\
J1042$+$0102 &0.116 &1010 &$2.38\pm 0.31$         &$42.67\pm 0.14$ &$43.66$ &6.46 &\nodata\\
J1214$+$0055 &0.396 &1980 &$2.70\pm 0.28$         &$43.93\pm 0.13$ &$44.50$ &7.62 &\nodata\\
J1259$+$0102 &0.394 &1460 &$0.25^{+0.80}_{-1.01}$ &$41.85\pm 0.45$ &$44.54$ &7.39 &1\\
J1311$+$0031 &0.429 &1640 &$3.06\pm 0.96$         &$43.27\pm 0.28$ &$44.70$ &7.60 &2\\
J1412$-$0035 &0.127 &1100 &$2.34\pm 0.38$         &$42.74\pm 0.14$ &$43.36$ &6.31 &\nodata\\
J1430$-$0011 &0.103 &1740 &$0.92\pm 0.64$         &$41.62\pm 0.47$ &$43.19$ &6.60 &\nodata\\
J1449$+$0022 &0.081 &1070 &$2.12\pm 0.16$         &$42.82\pm 0.08$ &$43.38$ &6.31 &3\\
J1451$-$0006 &0.139 &1120 &$2.56\pm 0.22$         &$42.96\pm 0.12$ &$43.57$ &6.48 &\nodata\\
J1705$+$6310 &0.119 &1660 &$2.40^{+1.02}_{-0.55}$ &$41.96\pm 0.25$ &$43.55$ &6.81 &1\\
J2338$+$0048 &0.170 &1010 &$2.69\pm 0.69$         &$42.86\pm 0.18$ &$43.63$ &6.43 &\nodata\\

\enddata

\tablenotetext{a}{Truncated to JHHMM$\pm$DDMM}
\tablenotetext{b}{Redshift from the SDSS catalog, as listed in WPM02}
\tablenotetext{c}{From WPM02}
\tablenotetext{d}{Derived from spectral fitting in {\tt Sherpa}, unless noted
otherwise.}
\tablenotetext{e}{Monochromatic, rest-frame 1~keV luminosity inferred from the
best--fit {\tt Sherpa} or HR$_{a/b}$ model flux, with quoted errors from 
{\tt Sherpa} or Poisson errors on the total number of counts respectively.}
\tablenotetext{f}{Monochromatic, rest-frame 5100\AA\ luminosity.
Fluxes are measured from
the original SDSS spectra used in WPM02; errors are 
considered negligible for $\mbh$ estimates.}
\tablenotetext{g}{All luminosities are calculated assuming
$H_0=70$\,km\,s$^{-1}$\,Mpc$^{-1}$, $\Omega_m=0.3$, $\Omega_\Lambda=0.7$.}
\tablenotetext{h}{Estimated from the \citet{kaspi00} relation.}

\tablecomments{
(1) $\Gamma$ derived from HR$_{\rm b}$;
(2) $\Gamma$ derived from HR$_{\rm a}$;
(3) Spectrum not well fit with power law $+$ Galactic absorption model; see
\S\ref{sec_771}.
}

\end{deluxetable}

\begin{deluxetable}{lccl}
\tablecolumns{4}
\tablewidth{250pt}
\tablecaption{Statistics of $\Gamma$ for NLS1 X-ray Samples \label{tab_gamma}}
\tablehead{
\colhead{Sample} &
\colhead{$\langle\Gamma\rangle$} &
\colhead{$\sigma_\Gamma$} &
\colhead{$\Gamma$ Range}
}
\startdata
BBF96 NLS1s &3.22 &0.73 &1.8--5.2\\
WPM02 (RASS) &2.75 &0.31 &1.5--4.4\\
This Paper &2.60 &0.39 &1.1--3.4\\
BBF96 Sy1s &2.31 &0.28 &1.4--3.4\\
\enddata
\end{deluxetable}

\end{document}